\documentclass[a4paper,twocolumn,english,aps,prb,amsmath,showpacs,amssymb]{revtex4}
\pdfoutput=1
\usepackage[T1]{fontenc}
\usepackage[latin9]{inputenc}
\usepackage{babel}
\usepackage{amsmath}
\usepackage{amssymb}
\usepackage{graphicx}
\usepackage{bm}
\usepackage{esint}
\usepackage{hyperref}



\begin{document}

\title{Scattering by linear defects in graphene: a continuum approach}

\author{J. N. B. Rodrigues$^{1,3}$, N. M. R. Peres$^{2}$ and J. M. B. Lopes
dos Santos$^{1}$}

\affiliation{$^{1}$CFP and Departamento de Física e Astronomia, Faculdade de
Ciências Universidade do Porto, P-4169-007 Porto, Portugal}

\affiliation{$^{2}$Centro de Física e Departamento de Física, Universidade do
Minho, P-4710-057, Braga, Portugal}

\affiliation{$^{3}$Graphene Research Centre, Faculty of Science, National University
of Singapore, 6 Science Drive 2, Singapore 117546}


\date{\today}
\begin{abstract}
We study the low-energy electronic transport across periodic extended defects
in graphene. In the continuum low-energy limit, such defects act as infinitesimally 
thin stripes separating two regions where Dirac Hamiltonian governs the low-energy 
phenomena. The behavior of these systems is defined by the boundary condition imposed 
by the defect on the massless Dirac fermions. We demonstrate how this low-energy 
boundary condition can be computed from the tight-binding model of the defect line. 
For simplicity we consider defect lines oriented along the zigzag direction, which
requires the consideration of only one copy of Dirac equation. Three defect lines of 
this kind are studied and shown to be mappable between them: the {\it pentagon-only}, 
the $zz(558)$ and the $zz(5757)$ defect lines. In addition, in this same limit, we 
calculate the conductance across such defect lines with size $L$, and find it to be 
proportional to $k_{F} L$ at low temperatures. 
\end{abstract}

\maketitle

\section{Introduction}


Graphene growth by chemical vapor deposition (CVD) on metal surfaces\cite{Li_Science:2009,Reina_NanoLett:2009,Kim_Nature:2009,Bae_NatureNanotech:2010}
is a very promising scalable method for producing graphene sheets.
However, the present status of the method, typically results in the synthesis of 
polycrystalline graphene abundant in topological defects, grain boundaries
(GBs) being, by far, the most common ones.\cite{Huang_Nature:2011,Kim_ACSNano:2011,Incze_APL:2011}

Due to graphene's hexagonal structure, pairs of pentagons and heptagons,
named Stone-Wales (SW) defects, as well as octagons, are expected
to form at graphene GBs.\cite{Stone_Science:1986} Recent atomic resolution
TEM studies \cite{Meyer_NL:2008,Lahiri_NatureNanotech:2010,Huang_Nature:2011,Kim_ACSNano:2011}
allowed the observation of GBs in CVD-grown graphene. These experimental
studies have shown that the GBs are generally not perfect straight
lines, and that the $5$-$7$ defects along the boundaries are not
periodic. Furthermore, as shown by recent TEM studies,\cite{Huang_Nature:2011,Kim_ACSNano:2011}
these extended pentagon-heptagon defect-lines intercept each other
at random angles, forming irregular polygons with edges showing a
stochastic distribution of lengths. This renders theoretical studies
of such defects difficult, in particular when using microscopic tight-binding
models.\cite{Ferreira_EPL:2011}

Theoretical studies have argued that GBs strongly influence the properties
of graphene, namely its chemical,\cite{Malola_PRB:2010} mechanical
\cite{Yazyev_PRB:2010,Grantab_Science:2011} and electronic ones.
Electronic mobilities of films produced through CVD are lower than
those reported on exfoliated graphene, because \cite{Novoselov_Science:2004,Novoselov_PNAS:2005}
electronic transport\cite{Neto_RMP:2009,Peres_RMP:2010} is hindered
by grains and GBs.\cite{Yazyev_NatureMat:2010,Ferreira_EPL:2011}

Recently, the observation of a linear extended defect acting as a
one-dimensional conducting charged wire\cite{Lahiri_NatureNanotech:2010}
stimulated some theoretical studies concentrated on the scattering
and transport properties of such wire.\cite{Bahamon_PRB:2011,Gunlycke_PRL:2011,Jiang_PLA:2011}
Most of these studies have so far been focused on tight-binding models.
\begin{figure}[!htp]
 \centering \includegraphics[width=0.9\columnwidth]{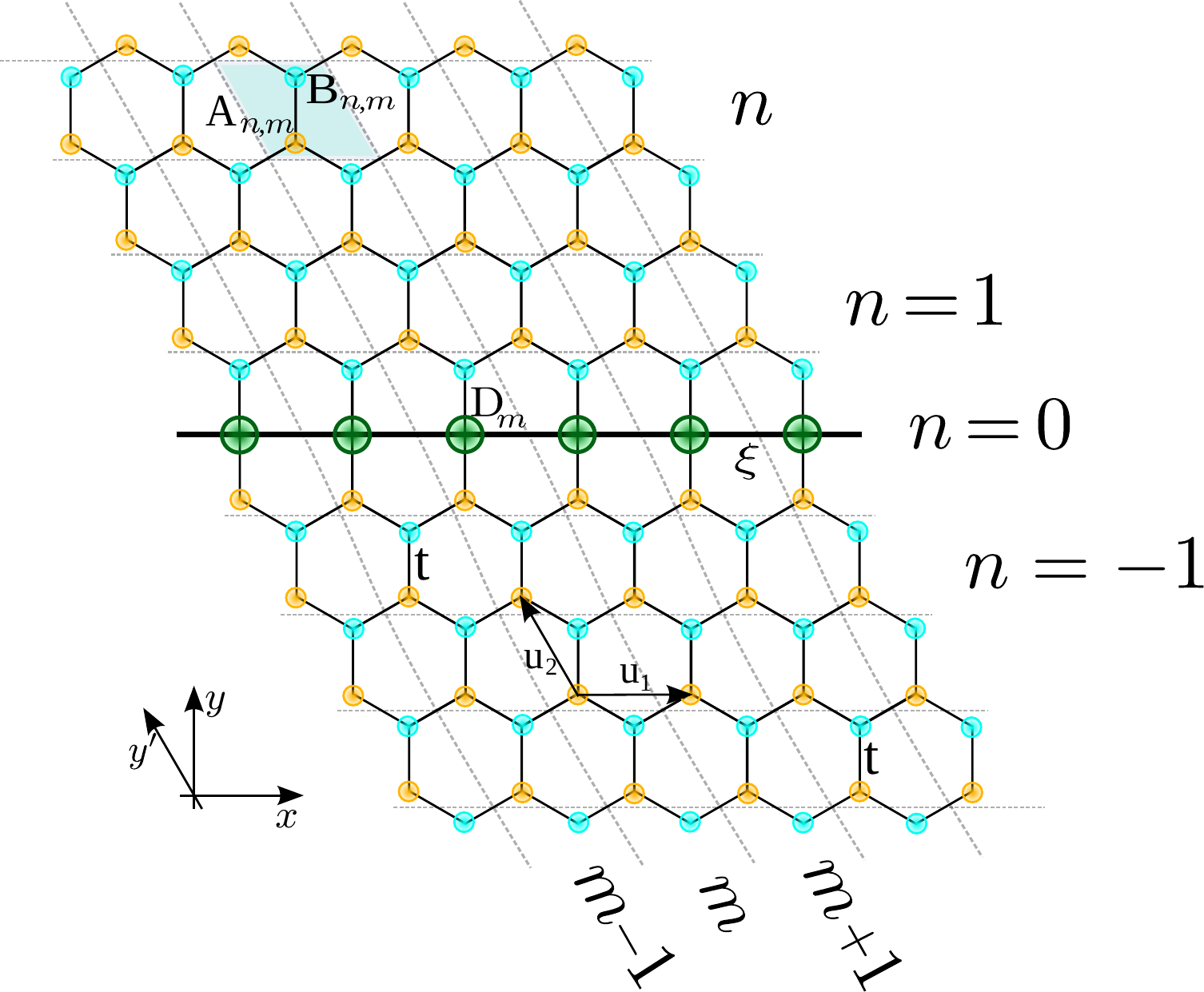}
\caption{(Color online) Scheme of a graphene sheet with a \textit{pentagon-only}
defect line along the zigzag direction. The primitive vectors are
$\mathbf{u}_{1}=a(1,0)$ and $\mathbf{u}_{2}=a(-1/2,\sqrt{3}/2)$.}
\label{fig:Scheme_GB_doubleJS} 
\end{figure}

The use of the continuum approximation on the scope of graphene have led to a better understanding
of many important phenomena occurring in graphene. Moreover, we believe that the use of this approach
in the study of the electronic scattering across extended defects in graphene, may further extend our 
insight onto the physics underlying these events in graphene.

As is widely known, a continuum approximation of graphene's first
neighbor TB Hamiltonian for states in the vicinity of the Dirac points,
describes graphene's low energy charge carriers as massless Dirac
fermions. These are governed by two copies of Dirac Hamiltonian, each
one of them valid around each of the Dirac points.\cite{Semenoff_PRL:1984}
In this continuum limit, the finite width defect line turns out to
essentially act as a one-dimensional (infinitesimally thin) line,
separating two distinct regions governed by Dirac Hamiltonian. The
defect line is modeled by a boundary condition on the Dirac spinors,
imposing a discontinuity across the defect. This boundary condition
determines the scattering properties of the defect.

For simplicity, in this text we only consider extended line defects oriented parallel
to the zigzag direction. In these cases, we can ignore intervalley scattering,
and thus consider only one copy of the Dirac equation. While some general
properties of the boundary condition and transmittance can be obtained
exclusively from the continuum description, the \emph{specific boundary
condition }must be derived from the TB model of the defect. Nevertheless,
we feel that this approach adds considerably to the understanding
of the low energy limit obtained from a TB description;\cite{Jiang_PLA:2011}
in particular, it explains, as we will show later, why different defects
can show exactly the same low energy transmittance.

To illustrate the main physical issues and the method of approach,
we start with a simplified version of a defect line, composed of a
double line of pentagons oriented along the zigzag direction of the
graphene lattice (see Fig. \ref{fig:Scheme_GB_doubleJS}), which we
dub as \textit{pentagon-only} defect line. Compared to the more realistic
linear defects that we treat later in the paper, the $zz(588)$ \cite{Lahiri_NatureNanotech:2010}
and the $zz(5757)$ defects, it has the added simplicity of full translation
symmetry along the defect direction, whereas the latter display a
doubling of the unit cell along that same direction. The low-energy
boundary conditions associated with these defect lines are also computed
and compared with that of the \textit{pentagon-only} defect line.
Suitable choices of the microscopic parameters lead exactly to the
same transmittance as a function of angle of incidence in all three
cases. Finally, we also compute the conductance across a defect of
length $L$ and find it to be proportional to $k_{F}L$ at low temperatures,
for all three defects considered.


\section{Electron transport across a pentagon-only grain boundary}


\subsection{The continuum description}

A graphene plane with an extended line defect can be viewed in the
low energy limit as two half-planes of massless Dirac Fermions, which
cannot be joined smoothly, because of the defect, a line of discontinuity.
To approach this problem, consider a finite strip of width $W$ in the
$y$ direction, where there is a general local potential, $\hat{V}(y)=V_{s}+V_{x}\sigma_{x}+V_{y}\sigma_{y}+V_{z}\sigma_{z}$,
for $\left|y\right|<W/2,$ such that $W\times(V_{s},\mathbf{V})\to(v_{s},\mathbf{v})$
as $W\to0$. Integrating the Dirac equation in the $y$ coordinate,
the resultant general boundary condition for the Dirac spinor has
the form (see Appendix~\ref{sec:The-general-low-energy-bc}) 
\begin{equation}
\Psi(x,0^{+})=\mathcal{M}\Psi(x,0^{-}),\label{eq:general_BC-1}
\end{equation}
 where the $2\times2$ matrix $\mathcal{M}$ reads 
\begin{equation}
\mathcal{M}=e^{-i\sigma_{y}(v_{s}+\mathbf{v}\cdot\boldsymbol{\sigma})/v_{F}},\label{eq:general_M-1}
\end{equation}
and $\boldsymbol{\sigma}=(\sigma_{x},\sigma_{y},\sigma_{z}).$ This
boundary condition has to satisfy the conservation of current in the
$y$ direction, \emph{i.e.},$\Psi^{\dagger}(x,0^{+})\sigma_{y}\Psi(x,0^{+})=\Psi^{\dagger}(x,0^{-})\sigma_{y}\Psi(x,0^{-})$
for any spinor, which implies that $\mathcal{M}^{\dagger}\sigma_{y}\mathcal{M}=\sigma_{y}$;
the form given in Eq.~\ref{eq:general_M-1} satisfies this condition.
An important feature, borne out by the derivation of Appendix~\ref{sec:The-general-low-energy-bc},
is energy independence of the boundary condition. When we integrate
the Dirac equation across the strip, and take the limit $W\to0$ ,
the term containing the energy $\epsilon$ of the state, which, unlike
the potential, is fixed, drops out. 

\begin{figure}[!htp]
 \centering \includegraphics[width=0.98\columnwidth]{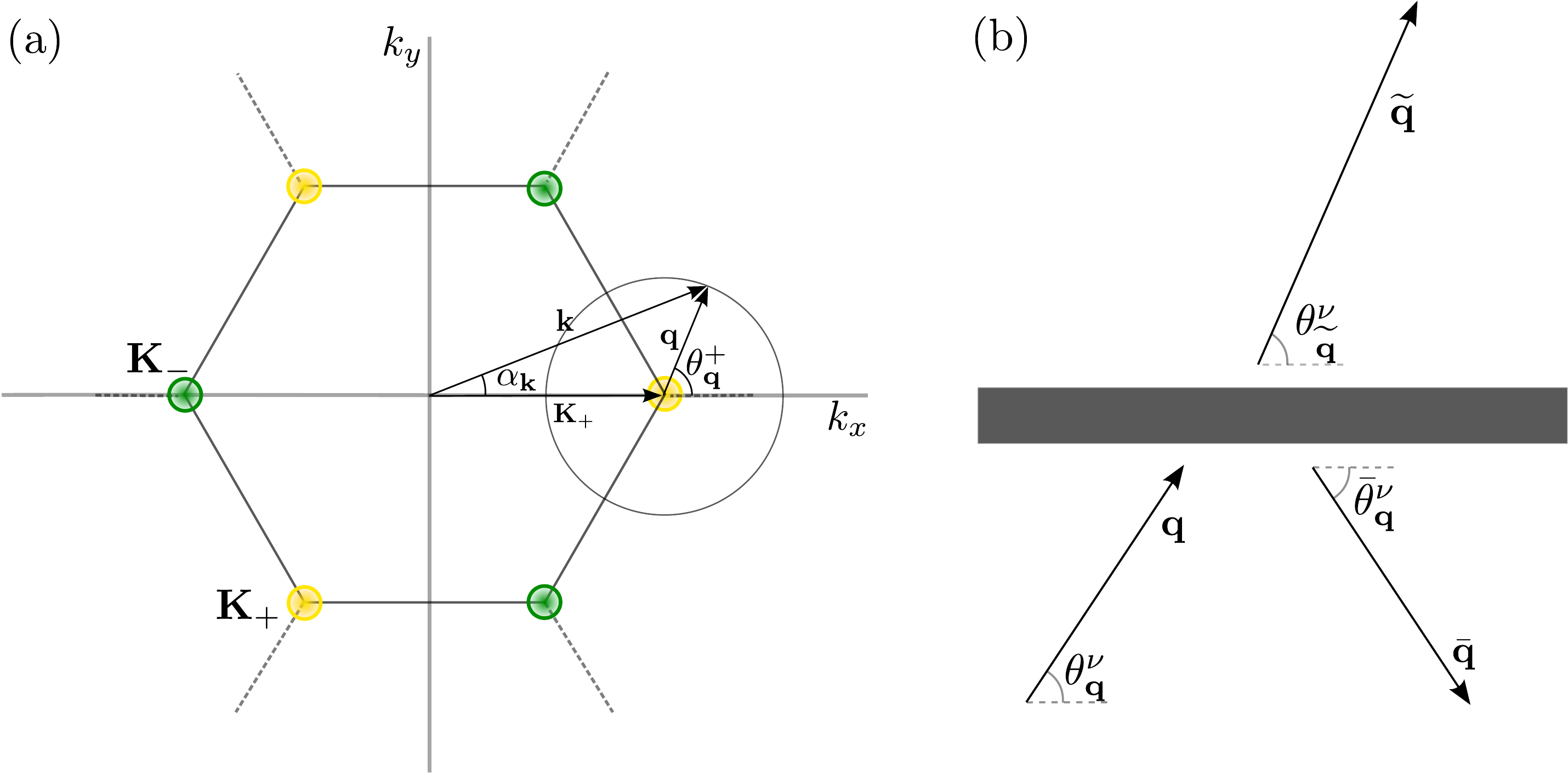}
\caption{(Color online) (a) Graphene FBZ with the incident vectors used in
the TB and CA formalism: $\mathbf{q}=\mathbf{k}-\mathbf{K_{\nu}}$.
(b) Scheme of the electron scattering through the barrier (in the
low-energy limit).}
\label{fig:FBZ} 
\end{figure}

An incoming wave from $y=-\infty$, will be partly reflected and partly
transmitted at the defect. As a consequence, the real-space wave-function
on each side of the defect line is given by 
\begin{subequations}
\label{eq:WaveFunctions_Dirac-1} 
\begin{eqnarray}
\Psi_{\mathbf{q} s}^{\nu}(\mathbf{r}) & = & \frac{1}{\sqrt{2}}\left[\begin{array}{c}
se^{-i\theta_{\mathbf{q}}^{\nu}}\\
1
\end{array}\right]e^{i(q_{x}x+q_{y}y)}\nonumber \\
 & + & \frac{\rho}{\sqrt{2}}\left[\begin{array}{c}
se^{-i\overline{\theta}_{\mathbf{q}}^{\nu}}\\
1
\end{array}\right]e^{i(q_{x}x-q_{y}y)},\quad y<0\\
\Psi_{\mathbf{q} s}^{\nu}(\mathbf{r}) & = & \frac{\tau}{\sqrt{2}}\left[\begin{array}{c}
se^{-i\theta_{\mathbf{q}}^{\nu}}\\
1
\end{array}\right]e^{i(q_{x}x+q_{y}y)},\quad y>0
\end{eqnarray}
\end{subequations}
where $\nu=\pm1$ specifies the Dirac cone, $\theta_{\mathbf{q}}$
is the complex phase of $\nu q_{x}+iq_{y}$, and $\overline{\theta}_{\mathbf{q}}^{\nu}=-\theta_{\mathbf{q}}^{\nu}$,
the complex phase of $\nu q_{x}-iq_{y}$ (see Fig. \ref{fig:FBZ}).
The sign of the energy is noted by $s$. Imposing the general boundary
condition gives immediately the following general expression for the
transmission probability
\begin{eqnarray}
T^{\nu}(E,\theta) & = & \frac{4\sin^{2}\theta}{\bigg|e^{i\nu2\theta}\mathcal{M}_{11}+\nu e^{i\nu\theta}\big(\mathcal{M}_{12}-\mathcal{M}_{21}\big)-\mathcal{M}_{22}\bigg|^{2}},\nonumber \\
\label{eq:trans-prob}
\end{eqnarray}
where we used the property $\left|\det\mathcal{M}\right|=1$, which
follows from the condition of flux conservation, $\mathcal{M}^{\dagger}\sigma_{y}\mathcal{M}=\sigma_{y}$.
A noteworthy feature, that follows naturally from this formulation,
is the energy independence of the transmission across the defect. 

To determine the\emph{ actual values }do the matrix elements of $\mathcal{M}$
for a specific defect in a graphene lattice we must consider its microscopic
description.

\subsection{The low energy limit of tight binding}


The first-neighbor TB Hamiltonian of graphene with a \textit{pentagon-only}
defect line (see Fig. \ref{fig:Scheme_GB_doubleJS}), can be written
as the sum of three terms, $\hat{H}=\hat{H}^{U}+\hat{H}^{D}+\hat{H}^{L}$,
where $\hat{H}^{U}$ ($\hat{H}^{L}$) stands for the Hamiltonian above
(below) the defect line, while the remaining term, $\hat{H}^{D}$,
describes the defect line itself. In second quantization the explicit
forms of $\hat{H}^{U}$ and $\hat{H}^{L}$ read 
\begin{eqnarray}
\hat{H}^{U(L)} & = & -t\sum_{m}\sum_{n}\bigg\{\Big[\hat{b}^{\dagger}(m,n)+\hat{b}^{\dagger}(m,n-1)\nonumber \\
 & + & \hat{b}^{\dagger}(m-1,n-1)\Big]\hat{a}(m,n)+h.c.\bigg\}\,,
\end{eqnarray}
 where for $H^{U}$ ($H^{L}$) $n\geq1$ ($n\leq-1$). The term describing
the defect, $H^{D}$, is 
\begin{eqnarray}
\hat{H}^{D} & = & -\sum_{m}\bigg\{\Big[\xi t\hat{d}^{\dagger}(m+1)+t\hat{a}^{\dagger}(m,0)\nonumber \\
 & + & t\hat{b}^{\dagger}(m,0)\Big]\hat{d}(m)+h.c.\bigg\},
\end{eqnarray}
 where $t$ is the usual hopping amplitude of pristine graphene and
$\xi t$ is the hopping amplitude between the $D_{m}$ atoms of the
defect line, as represented in Fig. \ref{fig:Scheme_GB_doubleJS}.

If we Fourier transform the Hamiltonian along the zigzag direction
($x$-direction), we reduce it to an effective one-dimensional chain
with two atoms per unit cell and a localized defect at its center
(see Fig. \ref{fig:Scheme_1D_doubleJSGB}). 
\begin{widetext}
\begin{center}
\begin{figure}[!htp]
 \includegraphics[width=0.85\textwidth]{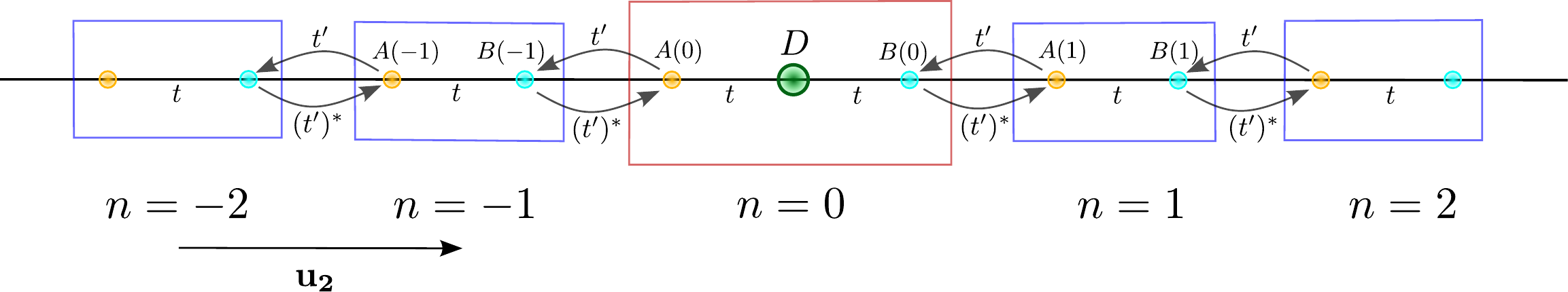}
\caption{(Color online) Scheme of the one-dimensional chain obtained by Fourier
transformation on the $x$-direction of the TB Hamiltonian of a graphene layer
with a \textit{pentagon-only} defect line along the zigzag direction. The
complex hopping amplitude $t'$ has the value $t'=t(1+e^{ik_{x}a})$.}

\label{fig:Scheme_1D_doubleJSGB} 
\end{figure}

\par\end{center}
\end{widetext}
The Hamiltonian of the effective chain is defined as 
\begin{equation}
\hat{H}(k_{x})=\hat{H}^{U}(k_{x})+\hat{H}^{D}(k_{x})+\hat{H}^{L}(k_{x})\,,\label{eq_HinK}
\end{equation}
 where the three terms on the right hand side of Eq.~(\ref{eq_HinK})
read 
\begin{subequations}
\label{eq:Hparts_1D} 
\begin{eqnarray}
\hat{H}^{U/L}(k_{x}) & = & -\sum_{n}\bigg\{\Big[t'\hat{b}^{\dagger}(k_{x},n-1)\nonumber \\
 & + & t\hat{b}^{\dagger}(k_{x},n)\Big]\hat{a}(k_{x},n)+h.c.\bigg\},\\
\hat{H}^{D}(k_{x}) & = & -2\xi t\cos(k_{x}a)\hat{d}^{\dagger}(k_{x})\hat{d}(k_{x})-\Big[t\hat{a}^{\dagger}(k_{x},0)\hat{d}(k_{x})\nonumber \\
 & + & t\hat{b}^{\dagger}(k_{x},0)\hat{d}(k_{x})+h.c.\Big].
\end{eqnarray}
 
\end{subequations}
The one-dimensional chain has alternating hopping amplitudes between
the atoms, $t$ and $t'=t(1+e^{ik_{x}a})$. Moreover, the electron
at a $D_{m}$ atom acquires an on-site energy term, $\widetilde{\epsilon}(k_{x})=-2\xi t\cos(k_{x}a)$,
which depends on the value of the longitudinal momentum $k_{x}$.

At the bulk of the one-dimensional chain ($n<-1$ and $n>0$), the TB equations for 
unit cell $n$ involve amplitudes at three different positions, $n-1$, $n$ and $n+1$, 
\begin{subequations} \label{eq:TBbulkEqs} 
\begin{eqnarray}
\epsilon A(k_{x},n) & = & -tB(k_{x},n)-(t')^{*}B(k_{x},n-1),\label{eq:TBbulkEqs1}\\
\epsilon B(k_{x},n) & = & -tA(k_{x},n)-t'A(k_{x},n+1).\label{eq:TBbulkEqs2}
\end{eqnarray}

\end{subequations}
Nevertheless, replacing $n\to n+1$ in Eq.~(\ref{eq:TBbulkEqs1}),
we can solve these equations for $A(k_{x},n+1)$ and $B(k_{x},n+1)$
and recast them as a recurrence relation relating amplitudes at unit
cell $n+1$ with those at unit cell $n$, 
\begin{eqnarray}
\mathbf{L}(n+1) & = & \mathbb{T}(\epsilon,k_{x}a).\mathbf{L}(n),\label{eq:BulkPassageMrel}
\end{eqnarray}
 where $\mathbf{L}(n)=[A(k_{x},n),B(k_{x},n)]^{T}$. The passage matrix,
$\mathbb{T}(\epsilon,\phi)$, is given by 
\begin{eqnarray}
\mathbb{T}(\epsilon,\phi) & = & -\frac{e^{-i\frac{\phi}{2}}}{2\cos\big(\frac{\phi}{2}\big)}\left[\begin{array}{cc}
1 & \frac{\epsilon}{t}\\
-\frac{\epsilon}{t} & 4\cos^{2}\big(\frac{\phi}{2}\big)-\frac{\epsilon^{2}}{t^{2}}
\end{array}\right].\label{eq:BulkPassageM}
\end{eqnarray}

The eigenvectors of matrix $\mathbb{T}(\epsilon,\phi)$ with eigenvalues 
with $\vert\lambda\vert^{2}=1$, correspond to Bloch solutions propagating
along the one-dimensional chain (band states). The
eigenvectors with eigenvalues $\vert\lambda\vert^{2}\neq1$
correspond to evanescent states which decrease when $n\to+\infty$
($n\to-\infty$) when $\vert\lambda\vert^{2}<1$ ($\vert\lambda\vert^{2}>1$).

Note that the previous formulation of the TB problem, is entirely
equivalent to the usual one, where translational symmetry along the
lattice vectors directions, allows the use of Bloch theorem to compute
the eigenvectors and eigenvalues of the TB Hamiltonian for pristine
graphene. 

A similar construction to that of Eq. (\ref{eq:BulkPassageMrel}) can be carried out in the rows 
containing the defect. The TB equations for the defect and its neighbors in the one-dimensional
chain are easily read from Fig.~\ref{fig:Scheme_1D_doubleJSGB}
\begin{subequations}
\label{eq_TB_Original} 
\begin{eqnarray}
\epsilon_{+}A(k_{x},1) & = & -(t')^{*}B(k_{x},0)-tB(k_{x},1), \label{eq_TB_Original-1} \\
\epsilon B(k_{x},0) & = & -tD(k_{x})-t'A(k_{x},1),\\
\epsilon D(k_{x}) & = & -t\big(A(k_{x},0)+B(k_{x},0)\big)\nonumber \\
 & - & 2\xi t\cos(k_{x}a)D(k_{x}),\\
\epsilon A(k_{x},0) & = & -(t')^{*}B(k_{x},-1)-tD(k_{x}),\\
\epsilon_{-}B(k_{x},-1) & = & -t'A(k_{x},0)-tA(k_{x},-1),
\end{eqnarray}
\end{subequations}
where $\epsilon_{\pm}=\epsilon\pm e\Delta V/2$, to account for a
possible potential difference between the two grains separated by
the \textit{pentagon-only} defect line. 
This set of TB equations can be used to construct a matrix equation
relating the TB amplitudes in opposite sides of the defect. The technique
is to solve each equation for the amplitude of the rightmost site
in Fig.~\ref{fig:Scheme_1D_doubleJSGB} and then cast them as $2\times2$
matrix equations. For instance, Eq.~(\ref{eq_TB_Original-1}) is equivalent
to
\begin{eqnarray}
\left[\begin{array}{c}
B(k_{x},1)\\
A(k_{x},1)
\end{array}\right] &=& \left[\begin{array}{cc}
-\frac{\epsilon_{+}}{t} & \frac{-(t')^{*}}{t}\\
1 & 0
\end{array}\right]\left[\begin{array}{c}
A(k_{x},1)\\
B(k_{x},0)
\end{array}\right].
\end{eqnarray}
With this procedure, one can derive 
\begin{eqnarray}
\mathbf{L}(1) & = & \mathbb{M}\mathbf{L}(-1),\label{eq:TB_BC}
\end{eqnarray}
where 
\begin{eqnarray}
  \mathbb{M} &\equiv& R M_{1}(\epsilon_{+},\phi) M_{2}(\epsilon,\phi) M_{3}(\epsilon,\phi) 
  M_{1}(\epsilon,\phi) M_{2}(\epsilon_{-},\phi) R^{T} \nonumber \\
\end{eqnarray}
is a $2\times2$ matrix; $R$ is the $\sigma_{x}$ Pauli matrix, used
to switch rows, $\phi=k_{x}a$, and $M_{1}$, $M_{2}$ and $M_{3}$
are 
\begin{subequations}
\label{eq:TB_BC_matrices} 
\begin{eqnarray}
M_{1}(\epsilon,\phi) & = & -\left[\begin{array}{cc}
\frac{\epsilon}{t} & (1+e^{-i\phi})\\
-1 & 0
\end{array}\right],\\
M_{2}(\epsilon,\phi) & = & -\frac{1}{1+e^{i\phi}}\left[\begin{array}{cc}
\frac{\epsilon}{t} & 1\\
-(1+e^{i\phi}) & 0
\end{array}\right],\\
M_{3}(\epsilon,\phi) & = & -\left[\begin{array}{cc}
\frac{\epsilon+2t\xi\cos(\phi)}{t} & 1\\
-1 & 0
\end{array}\right].
\end{eqnarray}
\end{subequations}
Note that the $2\times2$ boundary condition matrix, $\mathbb{M}$
{[}see Eq. (\ref{eq:TB_BC}){]}, depends on the energy, $\epsilon$,
on the longitudinal momentum, $k_{x}$, and on the potential difference
$\Delta V$, through $\epsilon_{+}$ and $\epsilon_{-}$.

We now have all the ingredients needed to compute the scattering coefficients
of an electron wave by the \textit{pentagon-only} defect line. Given
an incoming wave from $n=-\infty$, the presence of the defect (line)
at $n=0$, produces a reflected and a transmitted component. In such
a case, the wave-functions on each side of the defect (line) are given
by 
\begin{subequations}
\label{eq:WaveFunctions} 
\begin{eqnarray}
\mathbf{L}(n<0) & = & \lambda_{>}^{n}\mathbf{\Psi}_{>}+\rho\lambda_{<}^{n}\mathbf{\Psi}_{<},\\
\mathbf{L}(n>0) & = & \tau\lambda_{>}^{n}\mathbf{\Psi}_{>},
\end{eqnarray}
\end{subequations}
where $\rho$ and $\tau$ are, respectively, the reflection and transmission
scattering amplitudes, and $\mathbf{\Psi}_{>}$ and $\mathbf{\Psi}_{<}$
stand for the right and left moving eigenstates of matrix $\mathbb{T}(\epsilon,k_{x}a)$,
the passage matrix for pristine graphene, with corresponding eigenvalues
noted by $\lambda_{>}$ and $\lambda_{<}$. 
Imposing the boundary condition, Eq. (\ref{eq:TB_BC}), it is straightforward
to obtain the coefficients $\rho$ and $\tau$ for a given energy
and a given longitudinal momentum. In particular, $\tau$ reads 
\begin{eqnarray}
\tau & = & \frac{\det\widetilde{\mathbb{M}}}{\widetilde{\mathbb{M}}_{22}},\label{eq:TBtransPr}
\end{eqnarray}
\begin{subequations}
where $\widetilde{\mathbb{M}}=U^{-1}\mathbb{M}U$ is the boundary
condition matrix {[}see Eq. (\ref{eq:TB_BC}){]} in the eigenbasis
of the passage matrix of pristine graphene $\mathbb{T}(\epsilon,k_{x}a)$.
The transmission probability is given by $T=\left|\tau\right|^{2}=1/\left|\widetilde{\mathbb{M}}_{22}\right|^{2},$
since flux conservation again requires that $\left|\det\mathbb{M}\right|=1$.

But our main concern is the low energy limit. In the following we
assume $\Delta V=0$. Let us consider in parallel the equations that
propagate the state in the bulk and in the defect:
\begin{align}
\left[\begin{array}{c} A(k_{x},n+1) \\ B(k_{x},n+1) \end{array}\right] &= \mathbb{T}(\epsilon,\phi)
\left[\begin{array}{c} A(k_{x},n) \\ B(k_{x},n) \end{array}\right]\quad\textrm{bulk;}\label{eq:bulk_passage}\\
\left[\begin{array}{c} A(k_{x},1) \\ B(k_{x},1) \end{array}\right] & =\mathbb{M}(\epsilon,\phi)
\left[\begin{array}{c} A(k_{x},-1) \\ B(k_{x},-1) \end{array}\right]\,\textrm{defect;}\label{eq: defect_passage}
\end{align}
\end{subequations}
As is well known, near a Dirac point $\mathbf{K}_{\nu}$, the slowly
varying Dirac spinor $\Psi^{\nu}(\mathbf{r})$ is defined by (ignoring
irrelevant normalization constants)
\begin{eqnarray}
\Psi^{\nu}(m\mathbf{u}_{1}+n\mathbf{u}_{2}) &=& e^{-i\mathbf{K}_{\nu}\cdot
\left(m\mathbf{u}_{1}+n\mathbf{u}_{2}\right)}\left[\begin{array}{c} A(m,n)\\
B(m,n) \end{array}\right], 
\end{eqnarray}
and for a plane wave along $\mathbf{u}_{1}$
\begin{eqnarray}
  \Psi^{\nu}(m\mathbf{u}_{1}+n\mathbf{u}_{2}) &=& e^{-i\mathbf{K}_{\nu}\cdot n\mathbf{u}_{2}}\left[\begin{array}{c}
      A(k_{x},n) \\ B(k_{x},n) \end{array}\right]e^{i(\mathbf{k}-\mathbf{K}_{\nu})\cdot m\mathbf{u}_{1}} \nonumber \\
  &\equiv& \Psi^{\nu}(q_{x},n\mathbf{u}_{2})e^{i\mathbf{q}\cdot m\mathbf{u}_{1}}
\end{eqnarray}
where $\mathbf{q}=\mathbf{k}-\mathbf{K}_{\nu}$. This allows
us to recast Eqs.~(\ref{eq:bulk_passage}) and (\ref{eq: defect_passage})
in terms of the Dirac fields,
\begin{subequations}
\begin{align}
\Psi^{\nu}\left(q_{x},(n+1)\mathbf{u}_{2}\right) & =e^{-i\mathbf{K}_{\nu}\cdot\mathbf{u}_{2}}\mathbb{T}(\epsilon,\phi)\Psi^{\nu}(q_{x},n\mathbf{u}_{2}) , \label{eq:bulk-passage-1} \\
\Psi^{\nu}(q_{x},\mathbf{u}_{2}) & =e^{-i\mathbf{K}_{\nu}\cdot2\mathbf{u}_{2}}\mathbb{M}(\epsilon,\phi)\Psi^{\nu}(q_{x},-\mathbf{u}_{2}) , \label{eq:defect-passage-1}
\end{align}
\end{subequations}
where $\mathbf{K}_{\nu}\cdot\mathbf{u}_{2} = - \nu 2 \pi /3$.

If we take the Fourier transform with respect to the spatial variable
along $\mathbf{u}_{2}$ in Eq.~(\ref{eq:bulk-passage-1}), 
\begin{equation}
\Psi_{\mathbf{q}}^{\nu}=e^{i\nu2\pi/3}e^{-i\mathbf{q}\cdot\mathbf{u}_{2}}\mathbb{T}(\epsilon,\phi)\Psi_{\mathbf{q}}^{\nu}\label{eq:bulk-passage-2}
\end{equation}
In Appendix~\ref{sec:Dirac-equation-from-paasge} we show that the
matrix multiplying $\Psi_{\mathbf{q}}^{\nu}$ on the right hand side
tends to the identity matrix when $\mathbf{q},\,\epsilon\to0$; if
we expand the right hand side to linear order in $\epsilon$ and $\mathbf{q}$,
we obtain, as we should, the Dirac-Weyl equation (see Appendix~\ref{sec:Dirac-equation-from-paasge}).
However, at the defect, we find 
\begin{eqnarray}
e^{-i\nu2\pi/3}\mathbb{M}(\epsilon,\phi)\to\left(\begin{array}{cc}
0 & 1 \\ -1 & \xi \end{array}\right) \quad \textrm{when }\mbox{\ensuremath{\phi},\,\ensuremath{\epsilon\to}0},
\end{eqnarray}
which gives rise to the following equation 
\begin{eqnarray}
  \Psi^{\nu}(q_{x},\mathbf{u}_{2}) &=& \left(\begin{array}{cc} 0 & 1 \\ -1 & \xi \end{array}\right) 
  \Psi^{\nu}(q_{x},-\mathbf{u}_{2}) . \label{eq:BCmixC}
\end{eqnarray}

After Fourier transforming the previous equation in $q_{x}$, and as the continuum approximation yields
$a \to 0$ in $\mathbf{u}_{2}$, near the Dirac point, we end up concluding that the defect introduces a 
\emph{discontinuity} in the Dirac fields of the form we derived from general considerations,
$\Psi^{\nu}(x,0^{+})=\mathcal{M}\Psi^{\nu}(x,0^{-})$, with 
\begin{equation}
\mathcal{M}=\left(\begin{array}{cc}
0 & 1\\
-1 & \xi
\end{array}\right)\label{eq:passage-matrix-continuum-pent-pent}
\end{equation}
The transmission probability, given by the general expression of Eq.~(\ref{eq:trans-prob}),
becomes here
\begin{equation}
T^{\nu}(\theta)=\frac{\sin^{2}\theta}{1-\nu\xi\cos\theta+\xi^{2}/4}.\label{eq:trans-prob-pent-pent}
\end{equation}

\begin{figure}[!htp]
\centering \includegraphics[width=0.98\columnwidth]{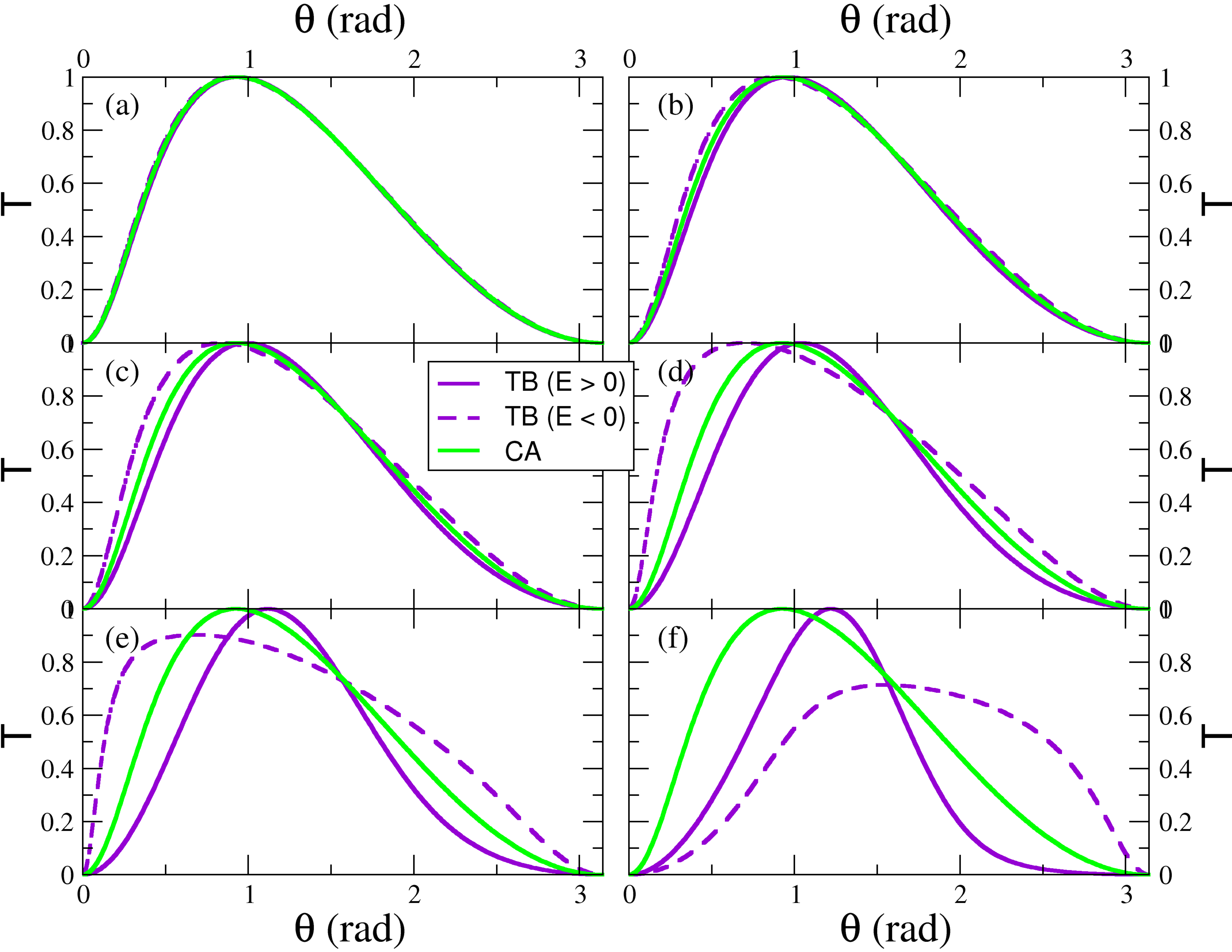}
\caption{(Color online) Plot of the transmission probability, $T$, in terms
of the angle of incidence on the \textit{pentagon-only} defect line,
$\theta_{\mathbf{q}}^{+}$ (for the low-energy limit, around $\mathbf{K}_{\nu}$,
with $\nu=+1$). We have used the value $\xi=1.2$ to obtain these
curves. In each of the panels, we compare the TB result for $\epsilon>0$
(full violet curves) and for $\epsilon<0$ (dashed violet curves),
with that obtained from the CA (in green), with $\Delta V=0$. Panel
(a), (b), (c), (d), (e) and (f), stand, respectively, for energies
$|\epsilon/t|=0.01$, $|\epsilon/t|=0.04$, $|\epsilon/t|=0.08$,
$|\epsilon/t|=0.16$, $|\epsilon/t|=0.32$ and $|\epsilon/t|=0.64$.}
\label{fig:Comparing_TB_CA} 
\end{figure}

In Fig. \ref{fig:Comparing_TB_CA}, we plot the transmission probability
$T$, in terms of the angle of incidence on the defect line, for both
the TB and the continuum approximation (CA), with $\xi=1.2$. The
various plots refer to different energies, but always to the same
Dirac point ($\nu=1$). As expected, the lower the energy, the better
the agreement between the TB and the CA results. For the other Dirac
point, the results are mirror-symmetric relatively to the normal incidence
angle $\theta=\pi/2$. 

A special case is of some interest, namely, for low energies, $\Delta V=0$,
and $\xi=2$, the transmission probability becomes $T^{\nu}(\theta)=(1+\nu\cos\theta)/2$,
in which case the \textit{pentagon-only} defect line acts as a valley
filter, for angles of incidence close to $\theta=0,\pi$. This same
feature has been found in another type of defect, the $zz(558)$,
which we consider in the next section, by Gunlycke and White;\cite{Gunlycke_PRL:2011}
this is no accident; we will show that these two defects share the
\emph{same }low energy limit.

It is worth noting, that since the passage matrix in the continuum
limit is obtained with $\phi=\mathbf{K}_{\nu}\cdot\mathbf{u}_{1}$
and $\epsilon=0$, it is easily got in a \emph{back of envelope }calculation,
by writing and solving the TB equations at zero energy. This procedure
is carried out in Appendix~\ref{app:Quick}.

It is expected that defect lines and grain boundaries in graphene
are reactive,\cite{Malola_PRB:2010} being a likely location for adsorption
of atoms or molecules. Such adsorbates, are expected to locally perturb
the properties of the defect lines. For simplicity, we may assume
that the adsorbate only modifies the local energy at the atom it adsorbs
to. We can account for such a phenomenon in the \textit{pentagon-only}
defect line, including in its TB model, an on-site energy, $\epsilon_{0}$
at the $D$ atoms of the defect line (see Fig. \ref{fig:Scheme_GB_doubleJS}
or Fig. \ref{fig:Scheme_1D_doubleJSGB}). Such a modification of the
TB model, will necessarily modify the TB boundary condition matrix,
$\mathbb{M}$ {[}see Eq.~(\ref{eq:TB_BC}){]}, as well as the continuum
approximation one, $\mathcal{M}$ {[}see Eq.~(\ref{eq:passage-matrix-continuum-pent-pent}){]}.
The TB boundary condition matrix, {[}see Eqs. (\ref{eq:TB_BC_matrices}){]},
will have its $\epsilon+2\xi t\cos(k_{x}a)$ term modified. This will
now include the on-site energy, $\epsilon_{0}$, as $\epsilon'=\epsilon+2\xi t\cos(k_{x}a)+\epsilon_{0}$.
In the CA limit, the boundary condition matrix, $\mathcal{M}$, will
have $(\xi t-\epsilon_{0})/t$ in the $\mathcal{M}_{22}$ entry of
the matrix instead of $\xi$. Thus, the adsorption of molecules at
the defect line, in very low energies, will be equivalent to rescaling
the hopping between the $D$ atoms at the defect line.


\section{The $zz(558)$ and the $zz(5757)$ defect lines}

\label{sec:zz558-zz5757}

\begin{figure}[!htp]
 \centering \includegraphics[width=0.98\columnwidth]{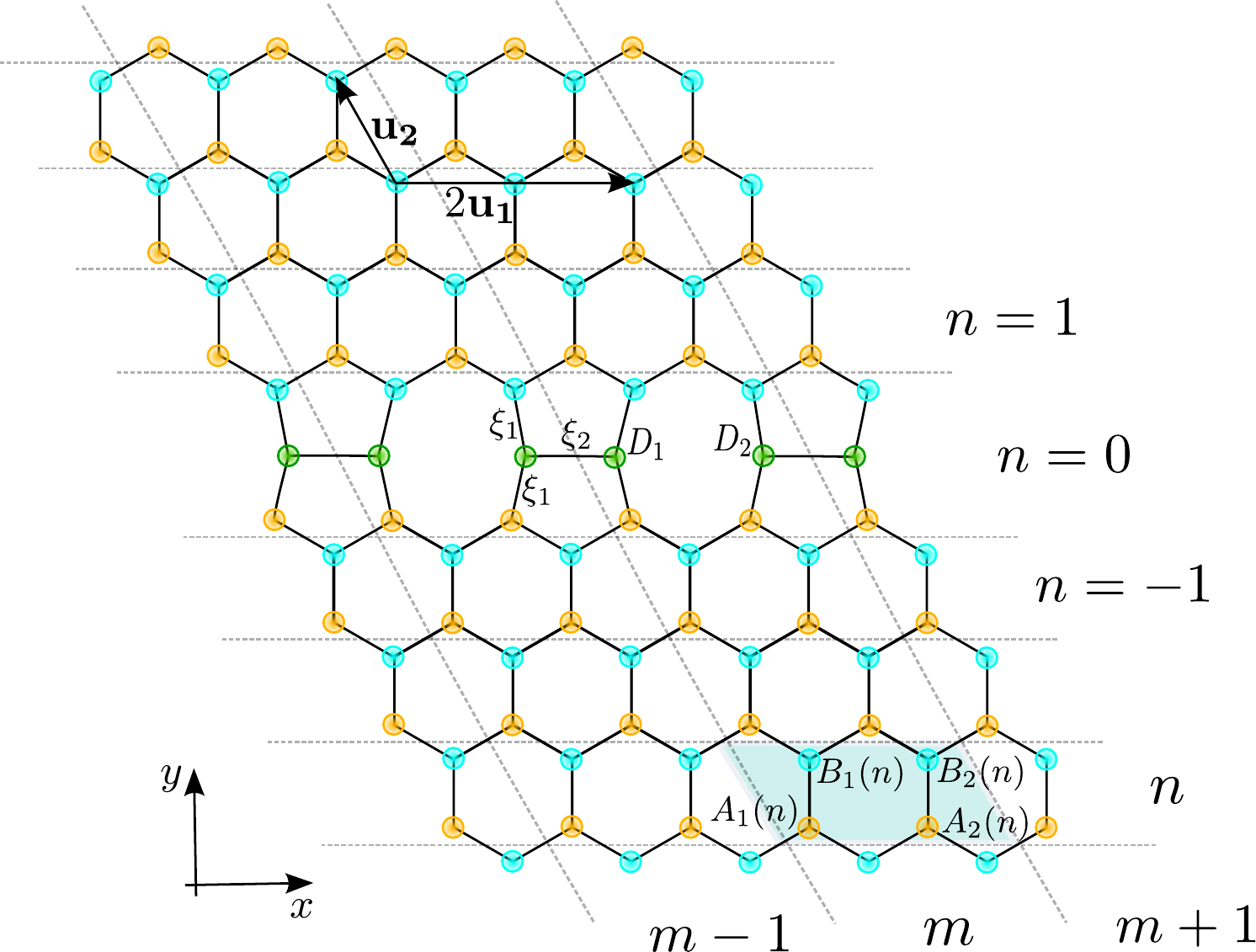}
\caption{(Color online) Scheme of a $zz(558)$ defect line.}

\label{fig:Scheme_zz558} 
\end{figure}

\begin{figure}[!htp]
\centering \includegraphics[width=0.98\columnwidth]{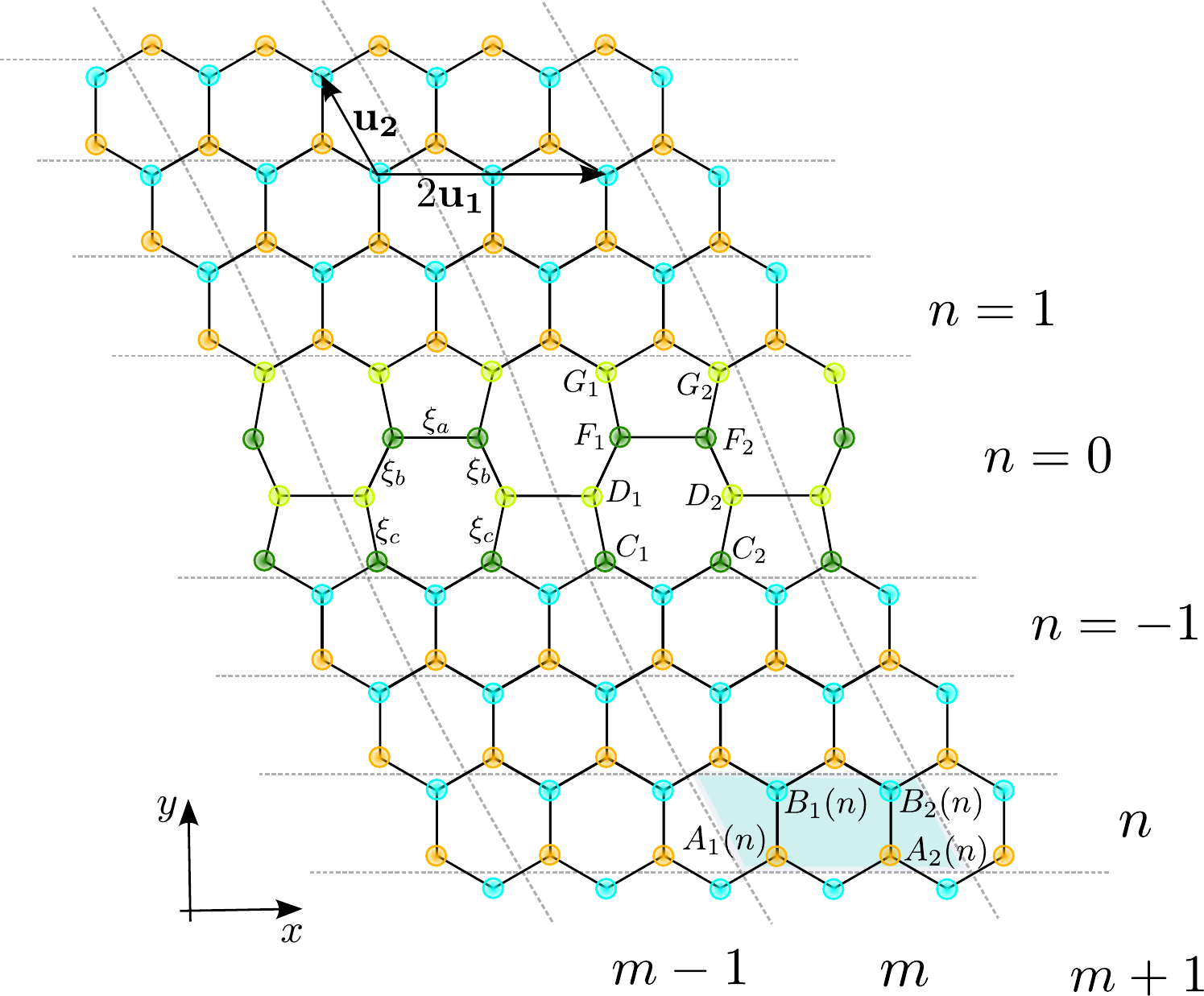}
\caption{(Color online) Scheme of a $zz(5757)$ defect line.}

\label{fig:Scheme_zz5757} 
\end{figure}

We now extend this treatment to the case of a $zz(558)$ defect line\cite{Lahiri_NatureNanotech:2010,Gunlycke_PRL:2011,Jiang_PLA:2011}
(see Fig. \ref{fig:Scheme_zz558}), and of a $zz(5757)$ defect line
(see Fig. \ref{fig:Scheme_zz5757}). 

We can proceed in close analogy with the case of a \textit{pentagon-only}
defect line treated in the previous section. But these more realistic
defects exhibit a feature that is not present in the previous case,
namely, the doubling of the unit cell in the direction parallel to
the defect. The corresponding folded First Brillouin Zone (FBZ) has
twice as many states at the same Bloch wave vector, as in the original
FBZ of graphene; the real space unit cell has two $A$ ($A_{1},\, A_{2}$)
and two $B$ ($B_{1},\, B_{2}$) sites. Around the new Dirac points,
now located at $\mathbf{K}_{\pm}=\pm\pi/3(1,-\sqrt{3})$, there will
be, in addition to two low-energy Dirac cones, two high energy bands.\cite{Rodrigues_PRB:2011}
At low energies, $\epsilon\approx0$, the extra states show up as
evanescent solutions.\cite{Ostaay_PRB:2011,Rodrigues_tobe:2012}

In pristine graphene we know the form of the high and low energy modes
since they are Bloch states of different wave vectors in the \emph{unfolded
}Brillouin zone. We can use this to define a change of basis that
decouples, in the bulk, these two energy sectors ($\phi=k_{x}a$):
\begin{eqnarray}
\left[\begin{array}{c}
A_{+}\\
B_{+}\\
A_{-}\\
B_{-}
\end{array}\right] &=& \Lambda(\phi)\left[\begin{array}{c}
A_{1}\\
B_{1}\\
A_{2}\\
B_{2}
\end{array}\right]
\end{eqnarray}
with 
\begin{eqnarray}
\Lambda(\phi) &:=& \frac{1}{\sqrt{2}}\left[\begin{array}{cccc}
1 & 0 & e^{-i\phi} & 0\\
0 & 1 & 0 & e^{-i\phi}\\
1 & 0 & -e^{-i\phi} & 0\\
0 & 1 & 0 & e^{-i\phi}
\end{array}\right].
\end{eqnarray}

Defining 
\begin{eqnarray}
\widetilde{\mathbf{L}}(n)\equiv[A_{+}(k_{x},n),B_{+}(k_{x},n),A_{-}(k_{x},n),B_{-}(k_{x},n)]^{T},
\nonumber \\
\end{eqnarray}
 we have, $\widetilde{\mathbf{L}}(n+1)=\mathbb{T}_{d}\widetilde{\mathbf{L}}(n)$,
where the matrix $\mathbb{T}_{d}$, written in blocks of $2\times2$
matrices, is
\begin{eqnarray}
\mathbb{T}_{\textrm{d}}(\epsilon,\phi) & = & \left[\begin{array}{cc}
\mathbb{T}_{+}(\epsilon,\phi) & 0\\
0 & \mathbb{T}_{-}(\epsilon,\phi)
\end{array}\right].\label{eq:doubledUCpassageM}
\end{eqnarray}
The $+$ and $-$ amplitudes propagate independently; $\mathbb{T}_{+}$
and $\mathbb{T}_{-}$, the passage matrices associated with the high
and the low-energy TB modes, are
\begin{subequations}
\label{eq:doubleUCpassageMss} 
\begin{eqnarray}
\mathbb{T}_{+}(\epsilon,\phi) & = & -\frac{e^{-i\frac{\phi}{2}}}{2\cos\big(\frac{\phi}{2}\big)}\left[\begin{array}{cc}
1 & \frac{\epsilon}{t}\\
-\frac{\epsilon}{t} & 4\cos^{2}\big(\frac{\phi}{2}\big)-\frac{\epsilon^{2}}{t^{2}}
\end{array}\right],\\
\mathbb{T}_{-}(\epsilon,\phi) & = & \frac{e^{-i\frac{\phi}{2}}}{2i\sin\big(\frac{\phi}{2}\big)}\left[\begin{array}{cc}
1 & \frac{\epsilon}{t}\\
-\frac{\epsilon}{t} & 4\sin^{2}\big(\frac{\phi}{2}\big)-\frac{\epsilon^{2}}{t^{2}}
\end{array}\right].
\end{eqnarray}
 \end{subequations}
The above computations are due to appear in a companion paper\cite{Rodrigues_tobe:2012}
devoted to the study of these same systems under the TB approach. 

In parallel with what we have done for the \textit{pentagon-only}
defect line {[}see Eqs. (\ref{eq_TB_Original})- (\ref{eq:TB_BC_matrices}){]},
using the TB equations at the $zz(558)$ or at the $zz(5757)$ defect
lines, it is possible to write an expression relating amplitudes at
the two sides of these defects, $\mathbf{L}(1)=\mathbb{M}\mathbf{L}(-1)$.
The matrix $\mathbb{M}$ is now a $4\times4$ matrix relating the
four amplitudes at each side of the defect line, and \textit{admix}\emph{ing},
in general, high and low-energy modes of different sides of the defect.
The high energy sector passage matrix in the bulk near $\epsilon=0$
and $\phi=\mathbf{K}_{\nu}\cdot\mathbf{u}_{1}=\nu\pi/3$ is 
\begin{equation}
\mathbb{T}_{+}\left(0,\nu\frac{\pi}{3}\right)=-e^{-i\pi/6}\left[\begin{array}{cc}
\frac{1}{\sqrt{3}} & 0\\
0 & \sqrt{3}
\end{array}\right];\label{eq:high-energy-passage}
\end{equation}
The corresponding eigenstates are evanescent, one growing exponentially
as $e^{(n\log3)/2}$, localized on the $B$ sub-lattice, and the other
decreasing as $e^{-(n\log3)/2}$, localized in the $A$ sub-lattice. 
This same result was obtained by Ostaay {\it et al.} in the scope of
the total reconstruction of the zigzag edge by Stone-Wales defects.\cite{Ostaay_PRB:2011}

\begin{figure}
  \centering
  \includegraphics[width=1\columnwidth]{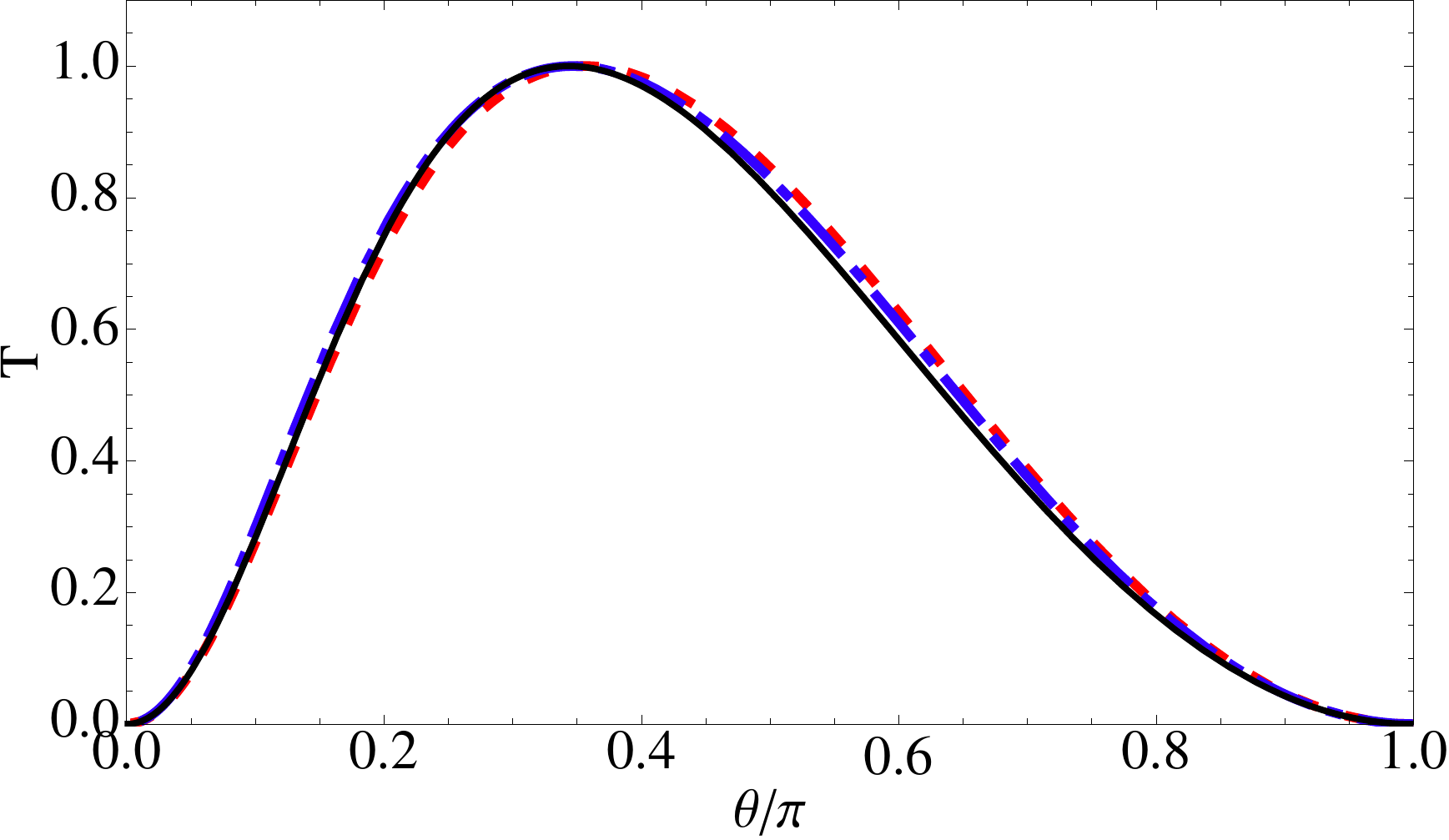}
  \caption{(Color online) The transmission probabilities for a
    $zz(558)$ defect, at $\epsilon/t=.03$, with
    $\xi_{1}=1,\,\xi_{2}=0.5$ (dashed red),
    $\xi_{1}=1.5,$~$\xi_{2}=1.125$ (dashed-dot, blue), obtained in a
    full TB calculation, and the corresponding low energy
    approximation (continuous black), given by
    Eq.~(\ref{eq:trans-prob-pent-pent}) with $\xi=1.0$.}
  \label{fig:three-plots}
\end{figure}

Given this, we conclude that a low energy state, must have the following form
in each one of the sides of the defect
\begin{subequations}
\begin{eqnarray}
\widetilde{\Phi}(k_{x},n) & \approx & \left[\begin{array}{c}
0\\
B_{+}(k_{x},n)\\
A_{-}(k_{x},n)\\
B_{-}(k_{x},n)
\end{array}\right]\qquad n<0;\label{eq:GenWFform-1}\\
\widetilde{\Phi}(k_{x},n) & \approx & \left[\begin{array}{c}
A_{+}(k_{x},n)\\
0\\
A_{-}(k_{x},n)\\
B_{-}(k_{x},n)
\end{array}\right]\qquad n>0.
\end{eqnarray}
\end{subequations}
This form fixes the $B_{+}(k_{x},-1)$ amplitude, in terms of the
low energy amplitudes $A_{-}(k_{x},-1)$ and $B_{-}(k_{x},-1)$, since
\begin{eqnarray}
\mathbb{M}_{22}B_{+}(k_{x},-1)+\mathbb{M}_{23}A_{-}(k_{x},-1)+\mathbb{M}_{24}B_{-}(k_{x},-1)=0,
\nonumber \\
\end{eqnarray}
and leads to an effective boundary condition relation for the low energy amplitudes
only. The latter reads
\begin{eqnarray}
\left[\begin{array}{c}
A_{-}(k_{x},1)\\
B_{-}(k_{x},1)
\end{array}\right]=\mathbb{M}^{\textrm{eff}}\left[\begin{array}{c}
A_{-}(k_{x},-1)\\
B_{-}(k_{x},-1)
\end{array}\right] ,
\end{eqnarray}
where the effective boundary condition matrix is obtained from matrix $\mathbb{M}$
\begin{eqnarray}
\mathbb{M}^{\textrm{eff}}=\left[\begin{array}{cc}
\mathbb{M}_{33}-\mathbb{M}_{32}\mathbb{M}_{23}/\mathbb{M}_{22} & \mathbb{M}_{34}-\mathbb{M}_{32}\mathbb{M}_{24}/\mathbb{M}_{22}\\
\mathbb{M}_{43}-\mathbb{M}_{42}\mathbb{M}_{23}/\mathbb{M}_{22} & \mathbb{M}_{44}-\mathbb{M}_{42}\mathbb{M}_{24}/\mathbb{M}_{22}
\end{array}\right], \nonumber \\
\end{eqnarray}
 The low energy sector, with the matrix $\mathbb{T}_{-}(\epsilon,\phi)$,
can be analyzed exactly as was done in Appendix~\ref{sec:Dirac-equation-from-paasge}
for the pentagon only boundary. We define the Dirac fields as before,
\begin{eqnarray}
\Psi^{\nu}(q_{x},n\mathbf{u}_{2})=e^{-i\mathbf{K}_{\nu}\cdot n\mathbf{u}_{2}}\left[\begin{array}{c}
A_{-}(k_{x},n)\\ B_{-}(k_{x},n) \end{array}\right] ,
\end{eqnarray}
so that, in the bulk 
\begin{eqnarray}
\Psi^{\nu}(q_{x},(n+1)\mathbf{u}_{2})=e^{-i\mathbf{K}_{\nu}\cdot\mathbf{u}_{2}}\mathbb{T}_{-}(\epsilon,\phi)\Psi^{\nu}(q_{x},n\mathbf{u}_{2}).  \nonumber \\
\end{eqnarray}
With a procedure entirely similar to the one detailed in Appendix~\ref{sec:Dirac-equation-from-paasge},
one finds, after Fourier transforming in $n$, that $\Psi_{\mathbf{q}}^{\nu}$
satisfies the Dirac equation. At the defect, 
\begin{eqnarray}
\Psi^{\nu}(q_{x},\mathbf{u}_{2})=e^{-i\mathbf{K}_{\nu}\cdot2\mathbf{u}_{2}}\mathbb{M}^{\mathrm{eff}}\left(0,\nu\frac{\pi}{3}\right)\Psi^{\nu}(q_{x},-\mathbf{u}_{2}) . \nonumber \\
\end{eqnarray}
The calculation of $\mathbb{M}^{\mathrm{eff}}$ yields 
\begin{equation}
\mathbb{M}^{\mathrm{eff}}\left(0,\nu\frac{\pi}{3}\right)=e^{i\nu2\pi/3}\left[\begin{array}{cc}
0 & 1\\
-1 & 2\frac{\xi_{2}}{\xi_{1}^{2}}
\end{array}\right],\label{eq:low-energy-passage-558}
\end{equation}
and so the boundary condition for the Dirac fields is 
\begin{equation}
\Psi^{\nu}(x,0^{+})=\left[\begin{array}{cc}
0 & 1\\
-1 & 2\frac{\xi_{2}}{\xi_{1}^{2}}
\end{array}\right]\Psi^{\nu}(x,0^{-}).\label{eq:low-energy-passage-558-2}
\end{equation}
It is remarkable that this has exactly the same form as found in the
pentagon--only boundary (\emph{c.f. }Eq.~{[}\ref{eq:passage-matrix-continuum-pent-pent}{]});
the low energy transmission probabilities of these two line defects
are \emph{the same }provided $2\xi_{2}/\xi_{1}^{2}=\xi.$ In Fig.~\ref{fig:three-plots}
we compare the transmission probabilities calculated with a full TB
calculation for different values of $\xi_{1}$ and $\xi_{2}$ but
the same value of $\xi_{2}/\xi_{1}^{2}$,\cite{Rodrigues_tobe:2012}
and the corresponding low energy approximation.

The treatment of the $zz(5757)$ line defect presents no further novelty.
Using the quick derivation method outlined in Appendix~\ref{app:Quick},
we arrive at the following passage matrix for the Dirac fields 
\begin{eqnarray}
  \mathcal{M}_{5757}^{\textrm{eff}}=\frac{-1}{2 \xi_{b}(\xi_{b}^{2}+\xi_{a}^{2}/2)}\left[\begin{array}{cc}
      a & b \\ -b & c \end{array}\right],\label{eq:EffBC_zz5757}
\end{eqnarray}
where $a = 2 \xi_{c}^{2}\big(\xi_{b}^{2}-\xi_{a}^{2}/4\big)$, $b=-\xi_{a}(\xi_{b}^{2}-\xi_{a}^{2})$
and $c=2(\xi_{b}^{4}+\xi_{a}^{4}+\xi_{b}^{2}\xi_{a}^{2})/\xi_{c}^{2}$ (see Fig~\ref{fig:Scheme_zz5757}
for the notation of the hopping amplitudes). The form of the passage
matrix (and the transmission probability) is not identical to the
previous cases, unless $\xi_{b}=-\xi_{a}/2$. 

\begin{figure}[!htp]
\centering \includegraphics[width=0.98\columnwidth]{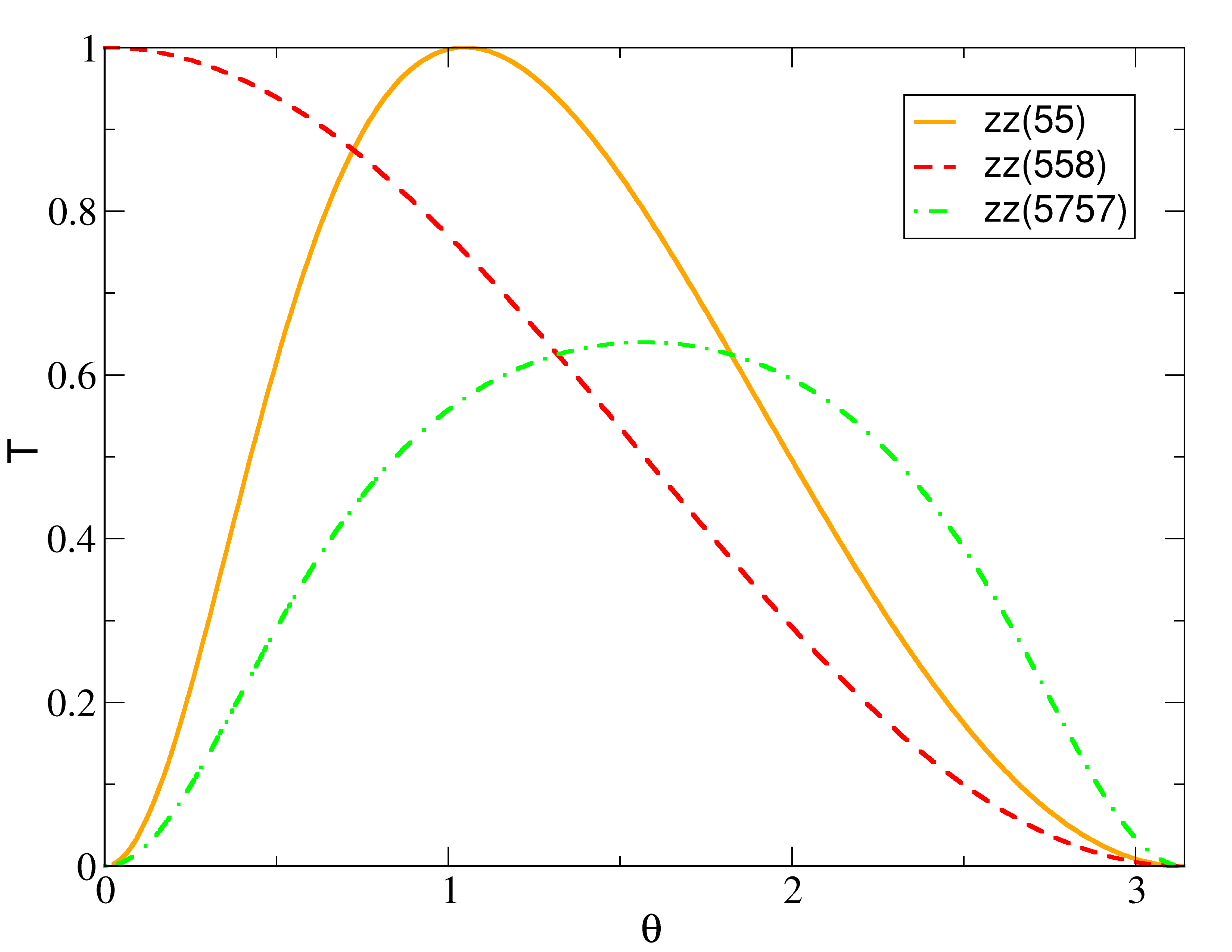}
\caption{(Color online) Comparison between the transmission probabilities,
$T$, across the defect for: the \textit{pentagon-only} {[}or $zz(55)${]}
defect line (full orange line); the $zz(558)$ defect line (dashed
red line); the $zz(5757)$ defect line (dot-dashed green line); The
transmission probabilities were calculated in the low energy limit
from Eq.~(\ref{eq:trans-prob}), with all the hopping parameters
were chosen equal to $1$, $\xi=\xi_{1}=\xi_{2}=\xi_{a}=\xi_{b}=\xi_{c}=1$.
$T$ is plotted in terms of the angle of incidence, $\theta_{\mathbf{q}}^{+}$,
of $\mathbf{q}=\mathbf{k}-\mathbf{K}_{+}$, the momentum around the
Dirac point $\mathbf{K}_{+}$. The transmission probabilities valid
for the vicinity of the other Dirac point, $\mathbf{K}_{-}$, are
obtained from the ones plotted above, by a reflection along the vertical
line $\theta=\pi/2$.}
\label{fig:Comparing-zz55-zz558-zz5757} 
\end{figure}

Despite the similarities, for general values of the hopping parameters,
the transmittances originating from each of the defect lines can be
considerably different. Such a case can be seen in Fig. \ref{fig:Comparing-zz55-zz558-zz5757},
where we compare the low-energy transmission probabilities associated
with each one of the previously discussed defect lines, for a special
situation with all hopping amplitudes equal to $t$, the bulk nearest
neighbor amplitude. 

\section{Conductance}

In this section we address the calculation of the linear conductance,
across a line defect, and show that the energy independence of the
transmission probability at low energies found in the previous three
cases gives rise to a conductance linear in $k_{F}$.

We make the usual assumption that the electron reservoirs at each
side of the defect line are in equilibrium and are thus described
by the single particle Fermi-Dirac distribution. Then the expression
for the total net current across the defect line, associated with
electrons living around the Dirac point $\mathbf{K}_{\nu}$, is given
by 
\begin{eqnarray}
J_{y}^{\nu}(\Delta V) & = & C\int_{-\infty}^{\infty}\textrm{d}E\left|E-\frac{e\Delta V}{2}\right|\mathbb{T}^{\nu}(E,\Delta V)\nonumber \\
 & \times & \left[f(E,\mu+\frac{e\Delta V}{2})-f(E,\mu-\frac{e\Delta V}{2})\right]\label{eq:CurrY}
\end{eqnarray}
 where $C=g_{s}e/(4\pi^{2}\hbar^{2}v_{F})$, $g_{s}$ stands for the
spin degeneracy, $\Delta V$ is the potential difference between the
each side of the defect line (see Fig. \ref{fig:Scheme_GB_doubleJS}),
and $f(E,\mu)$ the Fermi-Dirac distribution function for chemical
potential $\mu$; $\Delta\mu=e\Delta V$ is the difference between
chemical potentials at the two grains. In addition, $\mathbb{T}^{\nu}(E,\Delta V)$
stands for an angle-integrated transmission probability, 
\begin{eqnarray}
\mathbb{T}^{\nu}(E,\Delta V) & = & \int_{0}^{\pi}T^{\nu}(E,\Delta V,\theta)\sin\theta\textrm{d}\theta,
\end{eqnarray}
The total current is obtained summing the currents associated with
the two Dirac points $J_{y}=J_{y}^{+}+J_{y}^{-}$. One can verify
that $\mathbb{T}^{+}(E,\Delta V)=\mathbb{T}^{-}(E,\Delta V)\equiv\mathbb{T}(E,\Delta V)$,
and thus, $J_{y}=g_{v}J_{y}^{+}$, where $g_{v}=2$ is the valley
degeneracy.

\begin{figure}[!htp]
\centering \includegraphics[width=0.98\columnwidth]{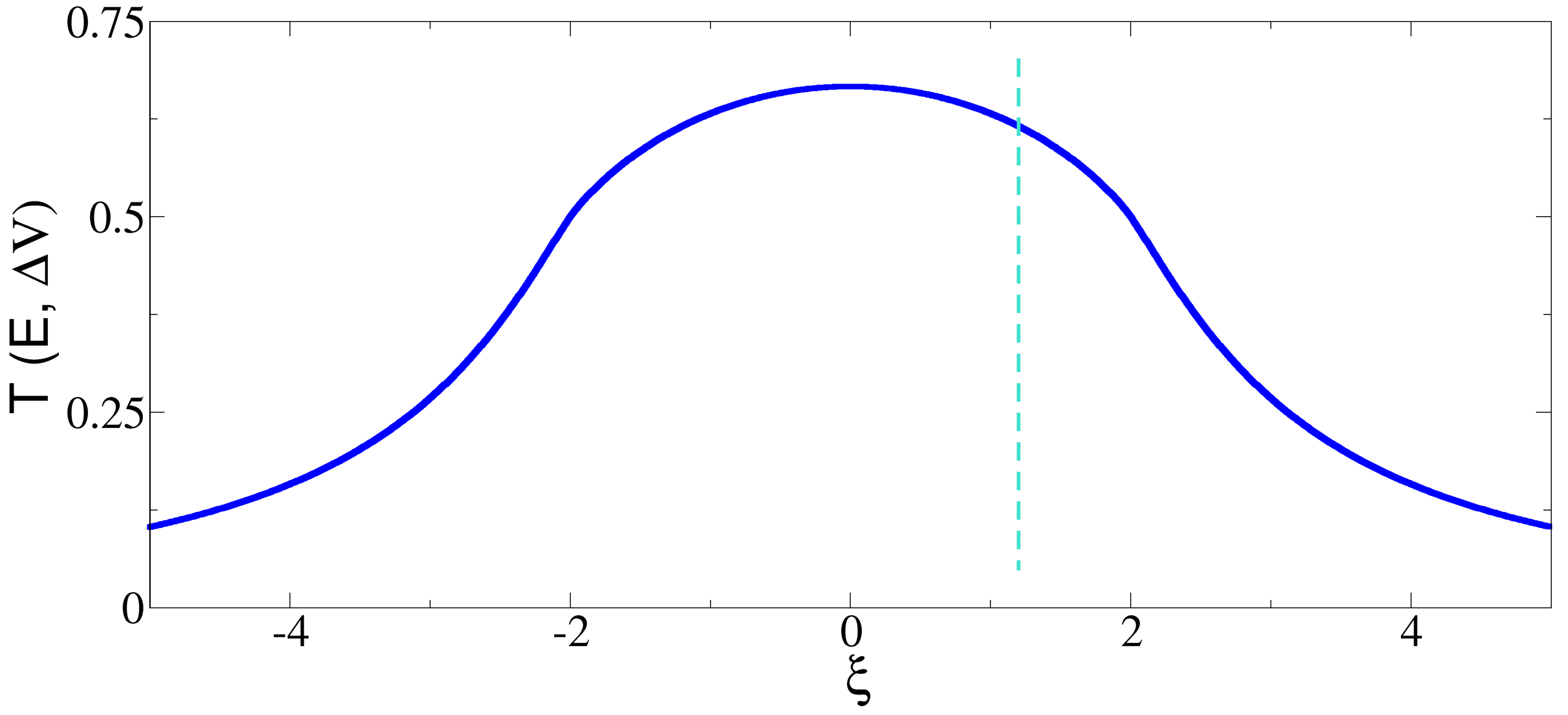}
\caption{(Color online) Dependence of the total angular integrated transmission
probability associated with one Dirac point, $\mathbb{T}(E,\Delta V)=\mathbb{T}^{+}(E,\Delta V)=\mathbb{T}^{-}(E,\Delta V)$,
when $\Delta V=0$, in terms of the hopping, $\xi t$, between the
atoms $D$ at the defect line (see Fig. \ref{fig:Scheme_GB_doubleJS}).
The vertical dashed line, indicates the value of $\xi$ used to obtain
the curves of Fig. \ref{fig:Comparing_TB_CA}: $\xi=1.2$.}

\label{fig:X_xi} 
\end{figure}

The conductance is then given by $G=LJ_{y}/\Delta V$, where $L$
is the length of the defect line, when the current is in the linear
regime, 
\begin{eqnarray}
G(T) & = & C'\int_{-\infty}^{\infty}|E|\mathbb{T}(E,0)\left(-\frac{\partial f(E,\mu)}{\partial E}\right)\textrm{d}E,\label{eq:LinPart_Cond}
\end{eqnarray}
 where $C'=Lg_{v}g_{s}e^{2}/(4\pi^{2}\hbar^{2}v_{F})$. The transmission
probability, $\mathbb{T}(E,\Delta V=0)=\mathbb{T}(0,0)$, does not
depend on $E$, as was seen above; it does depend on the values of
the hopping amplitudes in the vicinity of the defect, through the
passage matrix. {[}see Fig. \ref{fig:X_xi} , for the case of the
pentagon only defect{]}. We obtain, 
\begin{equation}
G(T)=Lg_{v}g_{s}\frac{e^{2}}{4\pi^{2}\hbar^{2}v_{F}}\mathbb{T}(0,0)\left[k_{B}T\times h\left(\frac{\mu}{k_{B}T}\right)\right],
\end{equation}
 where the function $h(x)$ is a Fermi integral, $h(x):=\int_{-\infty}^{+\infty}dy\left|y+x\right|\exp(y)/\left[\exp(y)+1\right]^{2}$,
with limits $h(0)=2\ln(2)$, and $h(x)\to\left|x\right|$ for $\left|x\right|\gg1$.
The conductance is linear in temperature for $\left|\mu\right|\ll k_{B}T$;
in the opposite limit, $\left|\mu\right|\gg k_{B}T$, it is practically
temperature independent, 
\begin{eqnarray}
G(T=0) & = & Lg_{v}g_{s}\frac{e^{2}}{4\pi^{2}\hbar^{2}v_{F}}\mathbb{T}(0,0)|\mu|\nonumber \\
 & = & g_{v}g_{s}\frac{e^{2}}{4\pi^{2}\hbar}\mathbb{T}(0,0)k_{F}L.
\end{eqnarray}

\section{Conclusion}

In this text we have focused on the study of the low-energy continuum limit behavior of the 
electronic transport across periodic defect lines oriented along the zigzag direction of 
graphene. We have argued that in this limit, such extended defects essentially act as 
one-dimensional infinitesimally thin lines, that separate two regions governed by the Dirac 
Hamiltonian. In the low-energy continuum limit the defective line imposes a 
boundary condition on the Dirac spinors at each side of the defect. It is this boundary 
condition that defines the low-energy behavior of the electronic transport of such systems.
We have demonstrated how can the boundary condition valid in the low-energy limit be computed
from the TB description of the defect line.

We have presented such a reasoning while working out the problem of the electronic transport across 
a {\it pentagon-only} defect line, finding its transmittance to be energy independent. Furthermore, 
we have also studied two other kinds of more realistic periodic defect lines: the $zz(558)$ defect 
line, recently observed in graphene sheets,\cite{Lahiri_NatureNanotech:2010} and the $zz(5757)$ 
defect line. We have briefly examined these latter cases, emphasizing the fact that their periodicity 
forces the appearance of high-energy modes at the Dirac points in addition to the typical low-energy 
Dirac modes. It has been shown that the influence of the former, can be encompassed in an effective 
boundary condition \textit{seen} by the low-energy massless Dirac fermions. The transmittance 
originating from such boundary conditions was again found to be energy independent. Furthermore, we 
have pointed out that the effective boundary conditions arising from the $zz(558)$ and $zz(5757)$ defect 
lines, turn out to be similar to the one arising from the \textit{pentagon-only} defect line. Moreover, 
the former can be mapped into the latter by an appropriate choice of the hopping parameters at the 
defects.

It is important to note that by expressing the transmission probability in terms of the boundary 
conditions satisfied by the Dirac fields at the defect, these results cast some light on the 
low-energy limit of the full TB calculations,\cite{Jiang_PLA:2011} leaving us with a better
understanding of the physics underlying such systems.

We have in addition shown how can we compute the low-energy limit conductance expression, across this 
kind of defect lines. Interestingly, at low temperatures, the conductance across a defect line of size 
$L$, turned out to be linear in $k_{F} L$. This feature originates from the energy independence of the 
low-energy transmittance of our defect lines.

Finally, we must mention, that the procedures presented in this text, can be used to solve more complex 
scattering problems in graphene. Not only linear and periodic defect lines, but also periodic curvilinear
extended defects oriented along graphene's zigzag direction, can be worked out using the framework 
presented in this text.

\begin{acknowledgements} J. N. B. R. was supported by Fundação para
a Ciência e a Tecnologia (FCT) through Grant No. SFRH/BD/44456/2008.
N. M. R. P. was supported by Fundos FEDER through the Programa Operacional
Factores de Competitividade - COMPETE and by FCT under project no.
PEst-C/FIS/UI0607/2011. J. M. B. L. S. was supported by Fundacão para
a Ciência e a Tecnologia (FCT) and is thankful for support and hospitality
of Boston University and of National University of Singapore. \end{acknowledgements}

\appendix

\section{The general low-energy boundary condition for a defect line oriented
along graphene's zigzag direction}

\label{sec:The-general-low-energy-bc}

In this appendix we derive Eq. (\ref{eq:general_M-1}), determining
the general form of the boundary condition matrix of a zigzag oriented
defect line in the continuum limit.

Suppose that our defect line is located in the region defined by $y\in[0,W]$.
Assume then, that in this region, we have a constant potential term
in the Dirac equation of the form 
\begin{eqnarray}
\hat{V} & = & V_{s}+V_{x}\sigma_{x}+V_{y}\sigma_{y}+V_{z}\sigma_{z}. \label{eq:GenPot}
\end{eqnarray}  
Our aim is to consider the limit where $W(V_{s},\mathbf{V})\to(u_{0},\mathbf{v}_{0})$
when $W\to0$ so that we can obtain the boundary condition of the
defect line in the continuum limit. We must refer that there are some works published on 
the literature, considering the electronic scattering across regions with potentials
that are a particular form of that given in Eq. \ref{eq:GenPot}.\cite{Katsnelson_Nature:2006,Gomes-Peres_JPCM:2008}

The Dirac equation in the region of the potential is 
\begin{eqnarray}
\epsilon\Psi_{\nu} & = & v_{F}\Big[\big(\nu\sigma_{x}(-i\partial_{x})+\sigma_{y}(-i\partial_{y})\big)+\big(V_{s}+\mathbf{V}\cdot\boldsymbol{\mathbf{\sigma}}\big)\Big]\Psi_{\nu}.\nonumber \\
\label{eq:DrcEqPotGen}
\end{eqnarray}
Since the defect line is oriented along the $x$-direction, we can
choose 
\begin{eqnarray}
\Psi_{\nu}(x,y) & = & \Phi_{\nu}(y)e^{iq_{x}x}
\end{eqnarray}
 which, after substitution into Eq. (\ref{eq:DrcEqPotGen}), results
in 
\begin{eqnarray}
\partial_{y}\Phi_{\nu}(y) & = & i\hat{P}\Phi_{\nu}(y),\label{eq:GenMintermdt1}
\end{eqnarray}
 where the operator $\hat{P}$ reads 
\begin{eqnarray}
\hat{P} & = & \frac{\sigma_{y}}{v_{F}}[\epsilon-\nu v_{F}q_{x}\sigma_{x}-V_{s}-\mathbf{V}\cdot\mathbf{\boldsymbol{\sigma}}]\label{eq:Poperator}
\end{eqnarray}
This first order differential equation (\ref{eq:GenMintermdt1}) can
be straightforwardly integrated, 
\begin{eqnarray}
\Phi_{\nu}(y) & = & e^{iy\hat{P}}\Phi_{\nu}(0).\label{eq:GenMintermdt2}
\end{eqnarray}
Taking now the limit $W(V_{s},\mathbf{V})\to(u_{0},\mathbf{v}_{0})$
when $W\to0$, we obtain the following expression for the continuum
limit of the boundary condition of a zigzag oriented defect line 
\begin{eqnarray}
\Phi_{\nu}(0^{+}) & = & e^{-i\sigma_{y}(u_{0}+\mathbf{v}_{0}\cdot\mathbf{\sigma})/v_{F}}\Phi_{\nu}(0^{-}),\label{eq:GeneralBCFinal}
\end{eqnarray}
 just as in Eqs. (\ref{eq:general_BC-1}) and (\ref{eq:general_M-1}).

As a final comment, we must stress the fact that the remaining terms
in $\hat{P}$, namely $\epsilon$ and $v_{F}q_{x}\sigma_{x}$, do
not contribute to the boundary condition when we take this limit;
they are fixed in value, unlike the potential terms, and cannot give
rise to a discontinuity when $W\to0$.

\section{Dirac equation from the passage matrix}

\label{sec:Dirac-equation-from-paasge}

In this appendix, we show how the Dirac-Weyl equation can be obtained
from the low-energy passage matrix relation in Eq.~(\ref{eq:bulk-passage-2})

\begin{eqnarray}
\Psi_{\mathbf{q}}^{\nu} & = & e^{i\nu2\pi/3}e^{-i\mathbf{q}\cdot\mathbf{u}_{2}}\mathbb{T}(\epsilon,\phi).\Psi_{\mathbf{q}}^{\nu}.\label{eq:bulk-passage-appendix}
\end{eqnarray}
Since $\phi=(\mathbf{K}_{\nu}+\mathbf{q})\cdot\mathbf{u}_{1}=\nu4\pi/3+\mathbf{q}\cdot\mathbf{u}_{1}$,
\begin{eqnarray}
\Psi_{\mathbf{q}}^{\nu} & = & e^{i\nu2\pi/3}e^{-i\mathbf{q}\cdot\mathbf{u}_{2}}\mathbb{T}\left(\epsilon,\nu\frac{4\pi}{3}+\mathbf{q}\cdot\mathbf{u}_{1}\right)\Psi_{\mathbf{q}}^{\nu}\label{eq:bulk-passage-appendix-1}
\end{eqnarray}
As we are working out a theory valid around the Dirac points, $\mathbf{q}$
and $\epsilon$ are small, and we can thus expand the exponential,
keeping solely the first order terms in the momentum,
\begin{widetext}
\begin{eqnarray}
  e^{i\nu2\pi/3} e^{-i\mathbf{q}\cdot\mathbf{u}_{2}} \mathbb{T}\left(\epsilon,\nu\frac{4\pi}{3}+\mathbf{q} 
    \cdot\mathbf{u}_{1}\right) &=& -\frac{e^{-i\mathbf{q}\cdot(\mathbf{u}_{1}/2+\mathbf{u}_{2})}}{2 \cos \big(
    \nu\frac{2\pi}{3}+\frac{\mathbf{q}\cdot\mathbf{u}_{1}}{2}\big)} \left[\begin{array}{cc}
      1 & \frac{\epsilon}{t} \\ -\frac{\epsilon}{t} & 4\cos^{2}\big(\nu\frac{2\pi}{3}+\frac{\mathbf{q}\cdot
        \mathbf{u}_{1}}{2}\big)-\frac{\epsilon^{2}}{t^{2}} \end{array}\right] \nonumber \\
  &\approx& I + \left[\begin{array}{cc} 0 & \frac{\epsilon}{t} \\ -\frac{\epsilon}{t} & 0 \end{array}
  \right] + \frac{\sqrt{3}}{2}a\left[\begin{array}{cc} -iq_{y}-\nu q_{x} & 0 \\ 0 & -iq_{y}+\nu q_{x}
    \end{array}\right]
\end{eqnarray}
\end{widetext}
where $I$ is the $2\times2$ identity matrix. This term cancels the one in
the right hand side of Eq.~(\ref{eq:bulk-passage-appendix-1}) and
we are left with
\begin{eqnarray}
\left[\begin{array}{cc}
0 & \frac{\epsilon}{t}\\
-\frac{\epsilon}{t} & 0
\end{array}\right]\Psi_{\mathbf{q}}^{\nu}+\frac{\sqrt{3}}{2}a\left[\begin{array}{cc}
-iq_{y}-\nu q_{x} & 0\\
0 & -iq_{y}+\nu q_{x}
\end{array}\right]\Psi_{\mathbf{q}}^{\nu}=0 . \nonumber \\
\end{eqnarray}
Upon multiplying by $i\sigma_{y}$, we obtain Dirac's equation, 
\begin{eqnarray}
\epsilon\Psi_{\mathbf{q}}^{\nu} & = & v_{F}\mathbf{\boldsymbol{\sigma}}_{\nu}\cdot\mathbf{q}\Psi_{\mathbf{q}}^{\nu},
\end{eqnarray}
 where $\mathbf{\boldsymbol{\sigma}}_{\nu}=(\nu\sigma_{x},\sigma_{y})$,
with the usual notation of $\nu=\pm1$ identifying the Dirac point.

\section{Quick derivation of the continuum limit boundary condition for the
\textit{pentagon-only} defect}

\label{app:Quick}

In this brief appendix, we will present a quick derivation of the
\textit{pentagon-only} defect low-energy boundary condition ($\epsilon=0$
and $q_{x}=0$), Eq.~(\ref{eq:passage-matrix-continuum-pent-pent}).

Let us start from the TB equations at the \textit{pentagon-only} defect,
Eqs. (\ref{eq_TB_Original}). In these equations, we begin by setting
$\epsilon=0$. In this way, the TB equations at the defect now read 
\begin{subequations} \label{eq_TB_Original-QuickMeth} 
  \begin{eqnarray}
    0 & = & -(t')^{*}B(k_{x},0)-tB(k_{x},1),\label{eq_TB_Original-QuickMeth-1}\\
    0 & = & -tD(k_{x})-t'A(k_{x},1),\label{eq_TB_Original-QuickMeth-2}\\
    0 & = & -t\big(A(k_{x},0)+B(k_{x},0)\big)\nonumber \\
    &  & -2\xi t\cos(k_{x}a)D(k_{x}),\label{eq_TB_Original-QuickMeth-3}\\
    0 & = & -(t')^{*}B(k_{x},-1)-tD(k_{x}),\label{eq_TB_Original-QuickMeth-4}\\
    0 & = & -t'A(k_{x},0)-tA(k_{x},-1),\label{eq_TB_Original-QuickMeth-5}
  \end{eqnarray}
\end{subequations}
where $t'=t(1+e^{-i k_{x}a/2})$. From now on, we set ourselves at $k_{x} a = \nu4\pi/3$.
The five equations written in Eqs. (\ref{eq_TB_Original-QuickMeth}), contain 7 amplitudes,
and we can solve them all in terms of $A(k_{x},-1)$ and $B(k_{x},-1)$;
we obtain for $A(k_{x},1)$ and $B(k_{x},1)$, 
\begin{eqnarray}
\left[\begin{array}{c}
A(k_{x},1)\\
B(k_{x},1)
\end{array}\right] & = & e^{i\nu\frac{2\pi}{3}}\left[\begin{array}{cc}
0 & 1\\
-1 & \xi
\end{array}\right]\left[\begin{array}{c}
A(k_{x},-1)\\
B(k_{x},-1)
\end{array}\right],\label{eq:PntOnlyBC-Quick}
\end{eqnarray}
which, using Eq.~(\ref{eq:defect-passage-1}), immediately identifies
the passage matrix for the Dirac fields, Eq.~(\ref{eq:passage-matrix-continuum-pent-pent}).
 
This procedure is quite general, and can be applied to the the other
line defects considered in this paper. In that case, however, we must
express the TB amplitudes in terms of the amplitude of the low and
high energy modes and set to zero the evanescent amplitudes of the
states that grow on each side of the defect. We can then solve for
the low energy amplitudes on one side of the defect, and obtain directly
the $2\times2$ passage matrix for the propagating modes.


\end{document}